\documentclass{article}
\usepackage{spconf,amsmath,graphicx}
\usepackage{booktabs}
\usepackage{amssymb}
\usepackage{url}
\usepackage[hidelinks]{hyperref}

\title{WPD++: An improved neural beamformer for simultaneous speech separation and dereverberation}
\name{Zhaoheng Ni$^{1,2,\star}$, Yong Xu$^2$, Meng Yu$^2$, Bo Wu$^3$, Shixiong Zhang$^2$, Dong Yu$^2$, Michael I Mandel$^1$
\thanks{$^{\star}$ This work was done while Z. Ni was a research intern at Tencent AI Lab, Bellevue, USA.}
}
\address{$^1$City University of New York, New York, NY, USA \\
$^2$Tencent AI Lab, Bellevue, WA, USA ~~~~
$^3$Tencent AI lab, Shenzhen, China}
\begin{document}
%
\maketitle
\begin{abstract}
This paper aims at eliminating the interfering speakers' speech, additive noise, and reverberation from the noisy multi-talker speech mixture that benefits automatic speech recognition (ASR) backend. While the recently proposed Weighted Power minimization Distortionless response (WPD) beamformer can perform separation and dereverberation simultaneously, the noise cancellation component still has the potential to progress. We propose an improved neural WPD beamformer called ``WPD++'' by an enhanced beamforming module in the conventional WPD and a multi-objective loss function for the joint training. The beamforming module is improved by utilizing the spatio-temporal correlation. A multi-objective loss, including the complex spectra domain scale-invariant signal-to-noise ratio (C-Si-SNR) and the magnitude domain mean square error (Mag-MSE), is properly designed to make multiple constraints on the enhanced speech and the desired power of the dry clean signal. 
Joint training is conducted to optimize the complex-valued mask estimator and the WPD++ beamformer in an end-to-end way.
The results show that the proposed WPD++ outperforms several state-of-the-art beamformers on the enhanced speech quality and word error rate (WER) of ASR. 
\end{abstract}
\begin{keywords}
Beamforming, speech separation, dereverberation, WPD, MVDR, multi-tap MVDR
\end{keywords}
\section{Introduction}
\label{sec:intro}

Solving the cocktail party problem \cite{haykin2005cocktail,wang2006computational} remains a challenging task due to the low signal-to-noise ratio of the signal, reverberation, and the presence of multiple talkers.
Recently, Neural Network (NN) based approaches show great potential in the speech separation task~\cite{luo2019conv, liu2020deep, zhang2020furcanext, luo2020dual, zeghidour2020wavesplit}. Those methods have high objective measure scores in terms of some objective metrics, however, it may inevitably introduce some non-linear speech distortion that downgrades the speech recognition performance \cite{du2014robust,xu2020neural}.
On the other hand, beamforming techniques \cite{benesty2011speech, benesty2008microphone}, e.g., minimum variance distortionless response (MVDR) \cite{habets2009new}, could extract the distortionless speech from the target direction.
Time-frequency (T-F) mask based beamforming approaches were successfully used for speech enhancement \cite{higuchi2016robust, heymann2016neural, heymann2017beamnet, erdogan2016improved, xiao2017time, xu2019joint, xu2020neural}. 

Simultaneous speech separation and dereverberation for the target speaker is the goal of this work. 
Weighted prediction error (WPE) \cite{nakatani2008blind, kinoshita2017neural} could remove the late reverberation. WPE followed by an MVDR beamformer was popularly used for speech separation, dereverberation, and ASR in the REVERB challenge~\cite{kinoshita2016summary} and the CHiME challenges~\cite{watanabe2020chime, boeddeker2020jointly}.
 Nakatani et al. \cite{nakatani2019maximum, nakatani2020jointly} unified the WPE and the weighted minimum power distortionless response (wMPDR) beamforming together into a single convolutional beamformer (WPD) for both speech dereverberation and enhancement. A mask-based WPD \cite{nakatani2020dnn} was proposed in a pipeline way where the T-F masks were estimated via a DNN, but the parameters of WPD were updated recursively. Zhang et al. \cite{zhang2020end} used the ASR loss to jointly optimize the real-valued mask estimator, WPD, and the acoustic model. However, the quality of the enhanced speech was not evaluated with the ASR loss only in \cite{zhang2020end}. Furthermore, the 
 generalization capability is always limited by the small far-field ASR dataset.

In this work, We propose an improved neural WPD beamformer method called ``WPD++'' that optimizes the neural network and the beamformer simultaneously. We jointly train the neural networks and WPD by utilizing the waveform level loss function. The enhanced speech is also evaluated on a general-purpose industry ASR engine to demonstrate the generalization capability of our enhancement model. Inspired by the multi-tap MVDR \cite{xu2020neural}, we improve the beamforming module in the conventional WPD by utilizing the spatio-temporal correlation to further strengthen the denoising capability of WPD. An additional novelty is that complex-valued masks, rather than the commonly used real-valued masks \cite{zhang2020end, nakatani2020dnn}, are estimated to calculate the covariance matrices of WPD++.


Another challenge we address is the loss function for the simultaneous speech separation and dereverberation.
Although the time domain Si-SNR \cite{luo2019conv} loss function could generate better performance for speech separation, it leads to worse performance for speech dereverberation \cite{luo2018real, tan2020audio}. One possible reason is that Si-SNR is too sensitive to the sample shift which is quite common in the convolutive reverberation. To alleviate this problem, we propose a multi-objective loss function to optimize the whole system in an end-to-end way. The multi-objective loss function includes magnitude domain mean square error (Mag-MSE) on the estimated dry clean power and a newly defined complex spectra domain Si-SNR (C-Si-SNR) on the final predicted waveform.

Our contributions in this paper are described in three parts. First, we propose a ``WPD++'' method where the spatio-temporal correlation is utilized to enhance the beamforming component of the conventional WPD. Secondly, we jointly train the complex-valued mask estimator and ``WPD++'' in an end-to-end way. The third contribution is that a multi-objective loss function is proposed to alleviate the limitation of the Si-SNR loss for the simultaneous speech separation and dereverberation.

The paper is organized as follows. In Sec. \ref{sec:wpd++}, the neural spatio-temporal MVDR and the proposed ``WPD++'' are illustrated. Sec. \ref{sec:loss} presents the introduced multi-objective loss function. Experimental setup and results are described in Sec. \ref{sec:exp}. Finally, conclusions are given in Sec. \ref{sec:conclusion}.


\section{Spatio-Temporal Neural Beamforming} \label{sec:wpd++}
Given a multi-channel speech mixture $y \in \mathbb{R}^{M\times N}$, where $M$ is the number of channels and $N$ is the number of the sampling points. The waveform signal $y$ can be transformed to the time-frequency signal $\bf{Y} \in \mathbb{C}^{M\times F \times T}$ by using Short Time Fourier Transform (STFT), where $F$ is the number of frequency bins and $T$ is the number of frames.  A beamformer aims at weighting sum the multi-channel signal into an enhanced signal $\bf{S} \in \mathbb{C}^{F\times T}$. The predicted signal $\hat{\bf{S}}(t,f)$ at frame $t$ and frequency bin $f$ can be modeled as:
\begin{equation}
\hat{\bf{S}}(t,f) = {\bf{w}^\mathsf{H}}(f){\bf{Y}}(t,f) 
\end{equation}
where ${\bf{w}} \in \mathbb{C}^{M \times F}$ is the weight matrix of the beamformer. 
\subsection{Complex-valued mask based spatio-temporal MVDR}
One solution to the MVDR beamformer which is based on reference channel selection \cite{subramanian2019investigation, habets2013two} is,
\begin{equation} \label{eq:mvdr_solution}
    \textbf{w}_{\text{MVDR}}(f) = \frac{{{\bf{\Phi}_{\textbf{NN}}^{-1}}(f) {\bf{\Phi}_{\textbf{SS}}}}(f)}{\text{Trace}({{{\bf{\Phi}_{\textbf{NN}}^{-1}}(f) \bf{\Phi}_{\textbf{SS}}}(f))}}\bf{u}
\end{equation}

where $\bf{\Phi}_{\textbf{NN}}$ and $\bf{\Phi}_{\textbf{SS}}$ are the covariance matrices of the noise and speech respectively. $\bf{u}$ is a one-hot vector representing the selected reference channel. Conventional mask-based MVDR applied the estimated real-valued ratio mask \cite{heymann2017beamnet, erdogan2016improved} to estimate $\bf{\Phi}_{\text{NN}}$ and $\bf{\Phi}_{\text{SS}}$. Here we estimate the complex-valued IRM (cIRM)~\cite{williamson2015complex} to boost the performance. cIRM is defined as
\begin{equation} \label{eq:cIRM}
\textbf{cIRM}  = \frac{\bf{Y}_r\bf{S}_r + \bf{Y}_i\bf{S}_i}{\bf{Y}_r^2 + \bf{Y}_i^2} + j*\frac{\bf{Y}_r\bf{S}_i-\bf{Y}_i\bf{S}_r}{\bf{Y}_r^2+\bf{Y}_i^2} = \frac{\bf{S}}{\bf{Y}}
\end{equation}
where the subscript $r$ and $i$ denote the real part and imaginary part of the STFT spectra respectively. Note that $\textbf{cIRM}$ is jointly trained in our framework by using the time domain loss, there is no need to do any scale compression. Then the estimated signal could be estimated as,
\begin{equation}
    \hat{\bf{S}}  = \textbf{cIRM} * \bf{Y} \\
     = (\bf{cIRM}_{r} + j*\textbf{cIRM}_{i} ) * (\bf{Y}_{r} + j*\bf{Y}_{i})
\end{equation}
where $\hat{\bf{S}} \in \mathbb{C}^{T \times F \times M}$ is the estimated multi-channel STFT for the target speech and \textbf{*} denotes the complex multiplication. The covariance matrix ${\bf{\Phi}_{\text{SS}}}$ of the target speech could be obtained as,
\begin{equation}
    {\bf{\Phi}_{\text{SS}}}(f) = \frac{\sum_{t=1}^{T}{\hat{\bf{S}}(t,f)\hat{\bf{S}}^\mathsf{H}(t,f)}}{\sum_{t=1}^{T}{\text{cIRM}^{\mathsf{H}}(t,f)\text{cIRM}(t,f)}}
\end{equation}

Xu et al. \cite{xu2020neural} further proposed a multi-tap MVDR method that estimates the covariance matrices by using the correlation of the neighbouring frames besides using the cross-channel correlation. The multi-tap expansion of the mixture is defined as $\overline{\bf{Y}}(t,f) = [\bf{Y}^\mathsf{T}(t,f), \bf{Y}^\mathsf{T}(t-1,f), ..., \bf{Y}^\mathsf{T}(t-L+1,f)]^\mathsf{T} \in \mathbb{C}^{ML \times 1}$. Note that the future taps would also be used if the system could be non-causal. 
The corresponding $\overline{\bf{S}}$, $\overline{\bf{N}}$ and $\overline{\textbf{cIRM}}$ could be defined in the same way.
Then the spatio-temporal covariance matrix of the target speech is calculated as
\begin{equation} \label{eq:mtmvdr_phi_ss}
    \Phi_{\bar{\bf{S}}\bar{\bf{S}}} = \frac{\sum_{t=1}^{T}{\overline{\bf{S}}(t,f)\overline{\bf{S}}^\mathsf{H}(t,f)}}{\sum_{t=1}^{T}{\text{cIRM}^\mathsf{H}(t,f)\text{cIRM}(t,f)}}
\end{equation}
The spatio-temporal covariance matrix of $\Phi_{\bar{\text{N}}\bar{\text{N}}}$ can be estimated in a similar way by replacing the speech mask $\textbf{cIRM}_s$ with the noise mask $\textbf{cIRM}_n$. Similar to Eq. (\ref{eq:mvdr_solution}), the multi-tap MVDR solution \cite{xu2020neural} is
\begin{equation}
{\overline{\textbf{w}}_{\text{MVDR}}}(f)=\frac{{{\bf{\Phi^{-1}_{\bar{\text{N}}\bar{\text{N}}}}}(f)}{\bf{\Phi_{\bar{\text{S}}\bar{\text{S}}}}}(f)}{\text{Trace}({{\bf{\Phi^{-1}_{\bar{\text{N}}\bar{\text{N}}}}}(f)}{{\bf{\Phi_{\bar{\text{S}}\bar{\text{S}}}}}(f)})}{\bar{\bf{u}}}~~~, ~~~~~{\bar{\bf{w}}}(f)\in \mathbb{C}^{ML \times 1}
\label{mtmvdr_solution}
\end{equation}
where $\bar{\bf{u}}$ is an expanded one-hot vector of $\bf{u}$ with padding zeros in the tail. The enhanced speech of the multi-tap MVDR \cite{xu2020neural} can be obtained as,
\begin{equation}
\hat{\bf{S}}(t,f)=\overline{\textbf{w}}^{\sf H}(f)\overline{\bf{Y}}(t,f)
\label{mtmvdr}
\end{equation}
However, the multi-tap MVDR in \cite{xu2020neural} was only designed and evaluated for the speech separation without dereverberation. In this work, simultaneous speech separation and dereverberation will be handled. Furthermore, we thoroughly investigate that the spatio-temporal correlation could be used to boost the performance of other beamformers, e.g., WPD.

\begin{figure*}[htb]
    \centering
    \includegraphics[width=\textwidth]{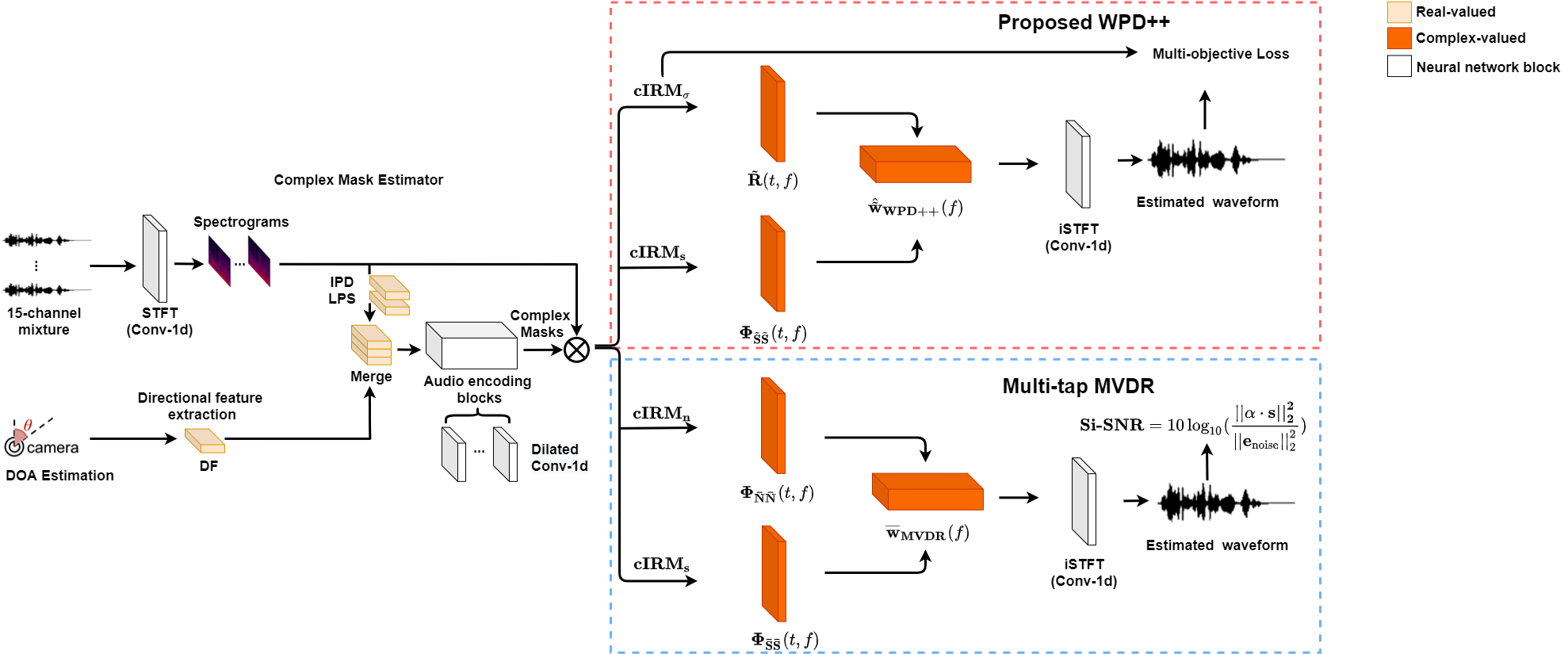}
    \caption{The overview of the proposed jointly trained complex-valued mask based WPD++ and the multi-tap MVDR systems.}
    \label{fig:system_overview}
\end{figure*}

\subsection{Proposed neural ``WPD++'' method}
The noisy speech can be decomposed into three parts:
\begin{equation}
    {\bf{Y}}(t,f) = {\bf{D}}(t,f) + {\bf{G}}(t,f) + {\bf{N}}(t,f),
\end{equation}
\begin{equation}
    {\bf{D}}(t,f) = \sum_{\tau=0}^{b-1}{{\bf{A}}({\tau},f){\bf{S}}({t-\tau},f)},
\end{equation}
\begin{equation}
    {\bf{G}}(t,f) = \sum_{\tau=b}^{L}{{\bf{A}}({\tau},f){\bf{S}}({t-\tau},f)},
\end{equation}
where $\bf{D}$ refers to the direct signal and early reflections, $\bf{G}$ refers to late reflections, and $\bf{N}$ refers to noises. $b$ is a frame index that could divide the reverberation into $\bf{D}$ and $\bf{G}$. If the desired signal is the direct path or the dry clean signal, then $b$ could be one. $\bf{A}$ denotes the acoustic transfer function. WPD \cite{nakatani2020jointly, nakatani2019maximum} aims at preserving the desired signal $\bf{D}$ while reducing $\bf{G}$ and $\bf{N}$.

The conventional WPD beamformer can be defined as
\begin{equation}
    {\hat{\overline{\textbf{w}}}}_{\text{WPD}}(f) = \frac{\textbf{R}^{-1}(f)\overline{\textbf{v}}(f)}{\overline{\textbf{v}}^H(f)\textbf{R}^{-1}(f)\overline{\textbf{v}}(f)}
\end{equation}
where $\overline{\bf{v}} = [\bf{v}, 0, 0, ..., 0]^T$ is the column vector containing the steering vector $\bf{v}$ and padding zeros. $\bf{R}$ is a spatio-temporal covariance matrix of the multi-tap multi-channel mixture signal $\bf{Y}$. $\bf{R}$ is weighted by the power of the target dry clean speech and defined as 
\begin{equation}
    \textbf{R}(f) =\sum_{t}{\frac{\overline{\bf{Y}}(t,f)\overline{\bf{Y}}^{\mathsf{H}}(t,f)}{{\bf{\sigma}^2}(t,f)}}
\end{equation}
where $\bf{\sigma}^2(t) = |\bf{D}^{(q)}(t)|^2$ is the time-varing power of the desired signal. $q$ denotes the reference microphone channel. Conventional WPD in \cite{nakatani2019maximum} iteratively estimate $\bf{\sigma}(t)$ and $\textbf{v}$. We apply a separate complex-valued mask for estimating $\bf{\sigma}$:
\begin{equation} \label{eq:sigma}
    \bf{\sigma} = |\textbf{cIRM}_\sigma * \bf{Y}^{(q)}|
\end{equation}

The steering vector $\bf(v)$ requires Eigenvalue decomposition which is not stable in neural network joint training~\cite{zhang2020end}. Zhang et al. \cite{zhang2020end} modified the original formula to avoid using the steering vector explicitly. 
\begin{equation}
{\hat{\overline{\textbf{w}}}}_{\text{WPD}}(f) = \frac{\textbf{R}^{-1}(f)({\bf{\Phi}_{\bar{\bf{S}}\bar{\bf{S}}}}(f))}{\text{Trace}({\textbf{R}^{-1}(f)({\bf{\Phi}_{\bar{\bf{S}}\bar{\bf{S}}}}(f))})}\overline{\bf{u}}
\end{equation}

The $\bf{\Phi}_{{\bar{\bf{S}}\overline{\bf{S}}}}$ is similar to the one (Eq. (\ref{eq:mtmvdr_phi_ss})) defined in the multi-tap MVDR beamformer. Normally the conventional WPE or WPD for the dereverberation would skip the neighbouring frames (a.k.a, a prediction delay) to avoid potential distortion on the speech of the current frame \cite{nakatani2008blind,nakatani2019maximum}. Given that the prediction delay exists, it only could estimate the desired signal with early reflections. However, the goal of our neural WPD++ model is to predict the direct path signal (a.k.a., dry clean) rather than the early reflections. On the other hand, neighbouring frames could benefit the beamforming for denoising and separation in spatio-temporal MVDR \cite{xu2020neural}, considering that the speech is highly correlated among neighbouring frames. The following WPD experiments with oracle mask in Sec. \ref{sec:exp} will show that neighbouring frames actually also help the wMPDR beamforming module in WPD. Furthermore, our proposed complex-valued mask based ``WPD++'' framework is jointly trained in an end-to-end way with the waveform level loss function. Hence the networks will automatically find the trade-off about how to use the neighbouring frames effectively. With the help of the highly correlated neighbouring frames, the ``WPD++'' beamforming weights are derived as:


\begin{equation}\label{eq:wpd++}
{\hat{\tilde{\textbf{w}}}}_{\text{WPD++}}(f) = \frac{\tilde{\bf{R}}^{-1}(f)({\bf{\Phi}_{\tilde{\bf{S}}\tilde{\bf{S}}}}(f))}{\text{Trace}({\tilde{\bf{R}}^{-1}(f)({\bf{\Phi}_{\tilde{\bf{S}}\tilde{\bf{S}}}}(f))})}\tilde{\textbf{u}}
\end{equation}

Different from the conventional WPD, we include the neighbouring frames in $\tilde{\bf{R}}$ and ${\bf{\Phi}_{\tilde{\bf{S}}\tilde{\bf{S}}}}$. Note that future neighbouring frames, which is also highly correlated with current frame, would be considered if the system could be non-causal. Another difference is that an utterance-level $\bf{\sigma}$-normalization is introduced to further normalize $\tilde{\textbf{R}}$,

\begin{equation}
    \tilde{\bf{R}}(f) =\frac
    {\sum_t(\frac{1}{{\bf{\sigma}^2}(t,f)}) {\tilde{\bf{Y}}({t,f})} {{\tilde{\bf{Y}}^{\mathsf{H}}({t,f})}}}
    {{\sum_t(\frac{1}{{\bf{\sigma}^2}(t,f)})}} 
\end{equation}
where $(\frac{1}{\bf{\sigma}^2(t,f)})$ could be regarded as a ``mask'' in the conventional mask based covariance matrix (e.g., Eq. \ref{eq:mtmvdr_phi_ss}). Intuitively, this ``mask'' would be larger with smaller $\bf{\sigma}$. It acts like a noise mask for the ``WPD++'' solution in Eq. (\ref{eq:wpd++}).



\section{Multi-objective loss function for neural ``WPD++'' joint training}\label{sec:loss}
Although Si-SNR \cite{luo2019conv} works well for speech separation, it leads to worse performance for speech dereverberation \cite{luo2018real, tan2020audio}. We design a multi-objective loss function for jointly training our proposed neural WPD++ model.
The Si-SNR \cite{luo2019conv} loss function is defined as
\begin{equation}
    \textbf{Si-SNR} = 10\log_{10}(\frac{||\alpha\cdot\bf{s}||^2_{2}}{||\textbf{e}_{\text{noise}}||^2_2})
\end{equation}
where $\alpha = \frac{<\hat{\bf{s}}, \bf{s}>}{||\bf{s}||_2^2}$, $\textbf{e}_{\text{noise}} = \hat{\bf{s}}- \alpha \cdot \bf{s} $, $\bf{s}$ and $\hat{\bf{s}}$ are the dry clean waveform and the estimated waveform respectively. 

The time-domain Si-SNR requires the estimated signal and the target signal are aligned perfectly. Thus it is very sensitive to the time-domain sample shift. However, the frame-level STFT might be less sensitive to the sample shift considering that the window size of STFT is always up to 512 samples for a 16kHz sample rate. Hence, we propose a complex-domain Si-SNR loss function that is less sensitive to the sample shift. Given the STFT of the estimation $\hat{\bf{S}}$ and the target reference $\bf{S}$, the function can be defined as:
\begin{equation}
    \textbf{C-Si-SNR} = 10\log_{10}(\frac{||\alpha\cdot\bf{S}||^2_{2}}{||\textbf{E}_{\text{noise}}||^2_2})
\end{equation}
\begin{equation}
    \alpha = \frac{<[\hat{\bf{S}}_r, \hat{\bf{S}}_i], [\bf{S}_r, \bf{S}_i]>}{||[\bf{S}_r, \bf{S}_i]||_2^2},
\end{equation}
\begin{equation}
    \textbf{E}_{\text{noise}} =  [\hat{\bf{S}}_r, \hat{\bf{S}}_i] - \alpha\cdot[\bf{S}_r, \bf{S}_i],
\end{equation}
where the real and imaginary components of $\bf{S}$ and $\hat{\bf{S}}$ are concatenated respectively for calculating C-Si-SNR. This guarantees the scale of the real and imaginary components are at the same level.

We also introduce the spectral MSE loss function which minimizes the difference between the estimated magnitude and the target magnitude. The spectral MSE loss is defined as:
\begin{equation} \label{eq:mag-mse}
    \textbf{Mag-MSE} =  \sum_t^T\sum_f^F{|| {\bf{S}}(t,f) - 
    {\hat{\bf{S}}}(t,f)||^2_2}
\end{equation}

As the accurate estimation of the magnitude of the desired signal $\bf{\sigma}$ (defined in Eq. (\ref{eq:sigma})) is the key to success of the WPD or WPD++ algorithm, a combo loss is designed for the prediction of $\bf{\sigma}$,
\begin{equation} \label{eq:combo-loss}
    \textbf{Combo-loss} =  \gamma\cdot\textbf{Mag-MSE}+\beta\cdot\textbf{Si-SNR}+\textbf{C-Si-SNR}
\end{equation}
where $\gamma$ and $\beta$ are used to weight the contribution among different losses. We empirically set $\gamma$ as 0.3 and $\beta$ as 1.0 to make the losses on the same scale. 
$\textbf{C-Si-SNR}$ loss only is used to optimize the final beamformed signal of WPD++.

\section{Experimental setup and results}
\label{sec:exp}

\subsection{Experimental setup and dataset}
\textbf{Dilated CNN-based Mask estimator:} We validate our proposed system and other methods on a multi-channel target speaker separation framework. Figure~\ref{fig:system_overview} describes the systems we use. The 15-element non-uniform linear microphone array is co-located with the 180 wide-angle camera. A rough Direction of Arrival (DOA) of the target speaker can be estimated from the location of the target speaker’s face in the whole camera view. We apply the location guided directional feature (DF) proposed by \cite{chen2018multi} that aims at calculating the cosine similarity between the target steering vector and the inter-channel phase difference (IPD) features. Besides the DF, we apply a $1\times 1$ Conv-1d CNN with the fixed STFT kernel to extract the Fourier Transform of the 15-channel speech mixture. Then we extract the log-power spectra (LPS) and interaural phase difference (IPD) features from the STFTs. The LPS, IPDs, and DF are merged and fed into a bunch of dilated 1D-CNNs to predict the complex-valued masks (as shown in Fig. \ref{fig:system_overview}). The 1-D dilated CNN based structure is similar to the ones used in the Conv-TasNet~\cite{luo2019conv}. The mask estimator structure is the same for all the methods. Before estimating the corresponding covariance matrices, we apply the spatio-temporal padding to the estimated STFT. Then we estimate the beamforming weights of MVDR, WPD, and WPD++ and finally obtain the estimated waveforms.

\textbf{Dataset:} The 200 hours clean Mandarin audio-visual dataset was collected from Youtube. The multi-channel signals are generated by convolving speech with room impulse responses (RIRs) simulated by the image-source method \cite{habets2006room}. The signal-to-interference ratio (SIR) is ranging from -6 to 6 dB. Also, noise with 18-30 dB SNR is added to all the multi-channel mixtures. The dataset is divided into 190000, 15000 and 500 multi-channel mixtures for training, validation, and testing. For the STFT conducted on the 16kHz waveform, we use 512 (32ms) as the Hann window size and 256 (16ms) as the hop size. The LPS is computed from the first channel of the noisy speech. In addition to the objective Perceptual Evaluation of Speech Quality (PESQ) \cite{rix2001perceptual} of the enhanced speech, we care more about whether the predicted speech could achieve a good ASR performance with an industry ASR engine for real applications. Hence, an Tencent industry general-purpose mandarin speech recognition API~\cite{tencent-api} is used to evaluate the word error rate (WER).

\textbf{Training hyper-parameters:} The networks are trained in a chunk-wise mode with a 4-second chunk size, using Adam optimizer with early stopping. The initial learning rate is set to 1e-3. Gradient clip norm with 10 is applied to stabilize the jointly trained MVDR \cite{xu2020neural}, multi-tap MVDR \cite{xu2020neural}, WPD \cite{nakatani2019maximum} and WPD++ (Proposed). PyTorch 1.1.0 is used.

\begin{table}[ht]
  \centering
  \begin{tabular}{l|l | c}
  \toprule
  Method   & Used frame tap(s) to calculate ${\bf{\Phi}}(t)$ & WER (\%)\\
  \hline
  \hline
  MVDR & t & 13.28 \\
  \hline
  MVDR & t-1, t & 11.90\\
  \hline
  MVDR & t-1, t, t+1 & 10.50\\
  \hline
  \hline
  WPD & t & 11.54\\
  \hline
  WPD & t-3, t & 10.22\\
  \hline
  WPD & t-4, t-3, t & 11.02 \\
  \hline
  \hline
  WPD++ & t-1, t & 10.50\\
  \hline
  WPD++ & t-1, t, t+1 & \textbf{9.48}\\
  \hline
  WPD++ & t-3, t-1, t, t+1 & 9.63\\
  \hline
  WPD++ & t-3[0:6], t-4[0:6], t-1, t, t+1 & 9.70 \\
  \bottomrule
\end{tabular}
\caption{The WER performances of several neural beamforming systems using the \textbf{oracle} cIRM masks. As for a reference, the WERs of the dry clean speech and the reverberant clean speech are 7.15\% and 8.26\%, respectively. There are 15 microphone channels for each frame.}
\label{tab:oracle-exp}
\end{table}

\subsection{Results and discussions}
\subsubsection{Evaluations for the spatio-temporal beamformers with \textbf{oracle} complex-valued masks}
To validate the capability of the proposed method, we firstly use the oracle target speech and noise cIRMs (i.e., calculated with oracle target speech and oracle noise in Eq. (\ref{eq:cIRM})) to compare the performances of different system settings. Table~\ref{tab:oracle-exp} shows the WER results of multi-tap MVDR, WPD, and the proposed WPD++ beamformers. 
Xu et al. \cite{xu2020neural} demonstrated that the neighbouring frames could improve the denoising performance of MVDR considering that the MVDR could use the spatio-temporal correlation. The experiments here with the oracle masks also prove that the performance of MVDR could be boosted by using neighbouring frames (even future frames) besides using the spatial cross-channel correlation. For example, the multi-tap MVDR could get 10.50\% WER which is lower than the 13.28\% WER of MVDR. 

Conventional WPD \cite{nakatani2019maximum} skips the neighbouring previous frames to predict the early reflections. As observed in Zhang et al.'s work \cite{zhang2020end}, WPD needs less previous frame taps when more microphones are available (15 linear non-uniform microphones are used in this work.). This is also aligned with our results that the WPD leads to worse performance when additional tap (i.e., $t-4$) is used.


Table~\ref{tab:oracle-exp} also shows the WPD++ beamforming achieves the best performance with $[t-1, t, t+1]$ frame taps. It demonstrates that the spatio-temporal correlation could also improve the performance of WPD. Note that our goal in this paper is to predict the direct path speech (or the dry clean speech), hence we can use the tap $t-1$. The future tap $t+1$ also helps to improve the performance of WPD++. This is because the future frame tap also highly correlates with the current frame $t$ given that the system could be non-causal. With the help of the spatio-temporal correlation, WPD++ could outperform the multi-tap MVDR \cite{xu2020neural} and the conventional WPD \cite{nakatani2019maximum} and obtain the lowest WER with 9.48\%. Additional temporal taps do not benefit the WPD++ model considering that $[t-1:t+1]$ taps have already been used with 15-channel for each frame. Another reason is that the Tencent ASR API \cite{tencent-api} is robust to some mild reverberation but not robust to interfering speech. The neighbouring frames could help more on the denoising function of the beamformer module in WPD++ considering that up to three competing speakers' speech might exist in our task.

\begin{table*}[htb]
\centering
\caption{PESQ and WER results for different purely NN-based or neural beamformer based speech separation and dereverberation systems using the predicted complex-valued masks across different scenarios.}
\label{tab:nn-exp}
\scalebox{0.79}{
\begin{tabular}{c|l|cccc|ccc|c|c}
\toprule
ID & \multicolumn{1}{c|}{Systems/Metrics} & \multicolumn{8}{c|}{PESQ $\in [-0.5,4.5]$~~~~~Reference=Dry Clean}   & WER ($\%$) \\ \hline
      &       & \multicolumn{4}{|c}{Angle between target and others} & \multicolumn{3}{|c|}{\# of speakers} &    &  \\
\hline
 & & 0-15$^{\circ}$ & 15-45$^{\circ}$ & 45-90$^{\circ}$ & 90-180$^{\circ}$ & 1spk & 2spk & 3spk & Avg. & Avg. \\ \hline
1 & Dry clean (reference) & 4.50 & 4.50 & 4.50 & 4.50  & 4.50 & 4.50 & 4.50 & 4.50 & 7.15\\
2 & Reverberant clean speech & 2.78  & 2.74 & 2.78 & 2.68 & 2.75 & 2.75 & 2.77 & 2.75 &  8.26\\
3 & Noisy Mixture (interfering speech + reverberation + noise) & 1.60 & 1.58 & 1.68 & 1.68 & 2.54 & 1.70 & 1.51 & 1.76 & 55.14 \\
\hline
4 & Purely NN with cIRM, Si-SNR loss & 2.22 & 2.36 & 2.52 & 2.44 & 3.01 & 2.43 & 2.25 &2.46 & 26.38 \\
5 & MVDR with cIRM \cite{xu2020neural} & 2.29 & 2.52 & 2.72 & 2.59 & 3.09 & 2.59 & 2.36 & 2.59 &  16.43\\ 
6 & Multi-tap MVDR with cIRM ([t-1:t+1] taps) \cite{xu2020neural} & 2.32 & 2.54 & 2.75 & 2.59 & 3.09 & 2.61 & 2.39 & 2.61 & 14.40 \\ 
\hline
7 & WPD ([t, t-3] tap) \cite{zhang2020end}, est: C-Si-SNR, $\sigma$: Combo-loss&   2.34 & 2.59 & 2.81 & 2.71 & \textbf{3.23} & 2.67 & 2.42 & 2.68 &  14.20 \\
\hline 
8 & Prop. WPD++, single mask, Si-SNR & 2.06 & 2.36 & 2.53 & 2.48 & 2.98 & 2.41 & 2.16 & 2.43 & 22.14 \\
9 & Prop. WPD++ , Si-SNR & 2.19 & 2.45 & 2.63 & 2.56 & 3.04 & 2.51 & 2.27 & 2.52 &  19.74 \\
10 & Prop. WPD++, C-Si-SNR & 2.19 & 2.45 & 2.67 & 2.61 & 3.09 & 2.53 & 2.27 & 2.54 & 18.78 \\
11 & Prop. WPD++, est: C-Si-SNR, $\sigma$: C-Si-SNR+Mag-MSE & 2.35 & 2.62 & 2.80 & 2.69 & 3.07 & 2.66 & 2.46 & 2.67 &13.86\\
12 & Prop. WPD++, est: Si-SNR, $\sigma$: Combo-loss & 2.36 & 2.62 & 2.81 & 2.70 & 3.09 & 2.67 & 2.46 & 2.67  & 13.51\\
13 & Prop. WPD++, same as 14, use  $\sigma$ 
as final est speech & 2.23 & 2.40 & 2.56 & 2.53 & 3.01 & 2.47 & 2.28 & 2.50 & 25.50 \\
14 & Prop. WPD++, est: C-Si-SNR, $\sigma$: Combo-loss & \textbf{2.47} & \textbf{2.71} & \textbf{2.89} & \textbf{2.76} & \text{3.18} & \textbf{2.76} & \textbf{2.55} & \textbf{2.76} & \textbf{12.04} \\
\bottomrule
\end{tabular}
}
\vspace{-6mm}
\end{table*}

\begin{figure}[ht]
    \centering
    \includegraphics[width=0.45\textwidth]{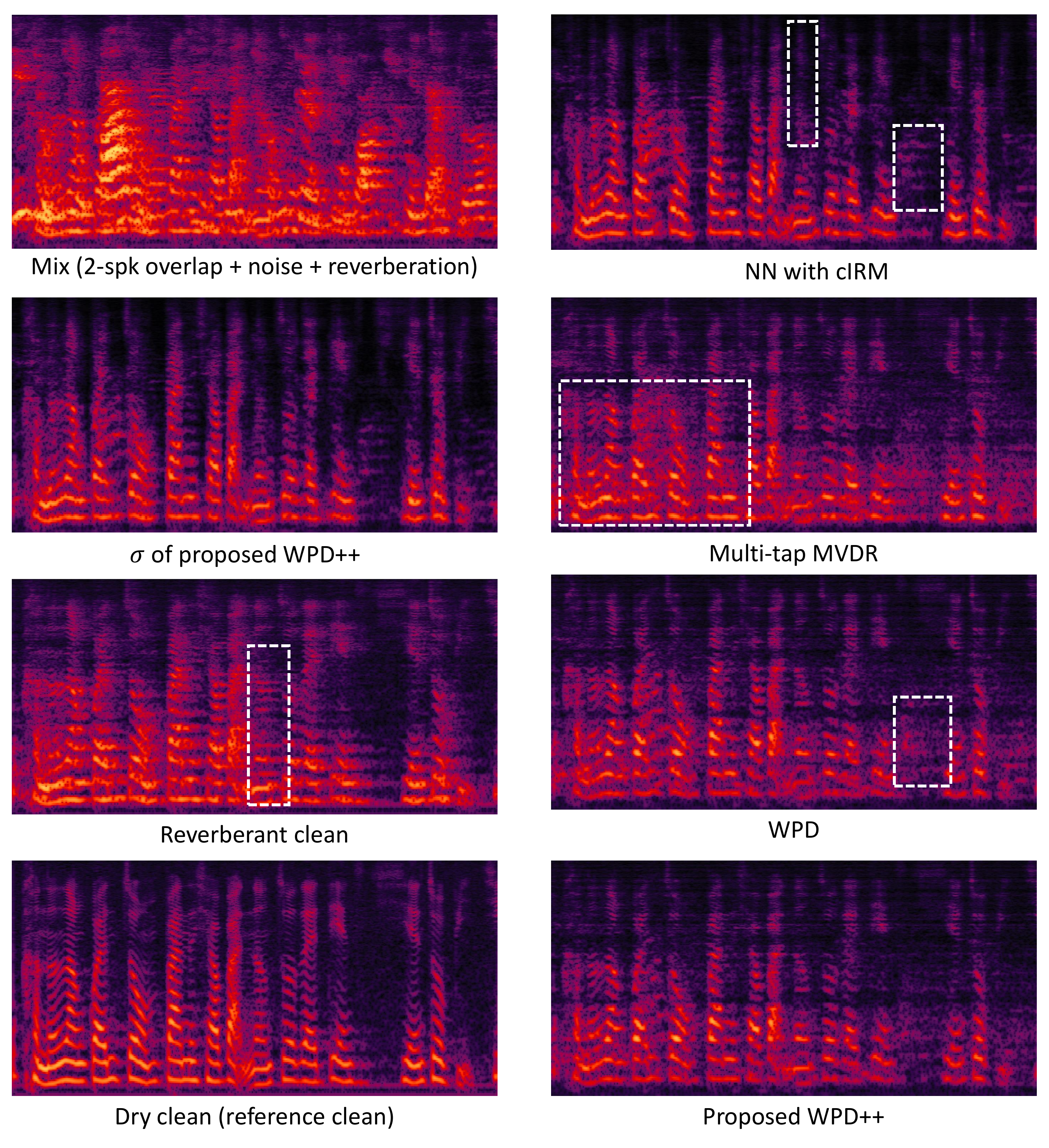}
    \caption{Spectrograms generated by different methods. 
    }
    \label{fig:spectral-demo}
\end{figure}

\subsubsection{Evaluations for neural beamformer systems with \textbf{predicted} complex-valued masks}

Based on the comparisons in the oracle experiments (shown in Table~\ref{tab:oracle-exp}), we choose the best temporal setting $[t-1, t, t+1]$ for the proposed neural WPD++ beamformer. 
Table~\ref{tab:nn-exp} shows that the proposed neural WPD++ beamformer (ID-14) with C-Si-SNR loss for the estimated signal (denoted as ``est'') and the combo-loss for $\sigma$ (i.e., the magnitude of the dry clean speech which is defined in Eq. (\ref{eq:sigma}).) achieves the best PESQ (2.76 on average) and lowest WER (12.04\%). Compared to the best multi-tap MVDR system (ID-6) and the best conventional WPD system (ID-7), the proposed neural WPD++ method (ID-14) could get relative 16.4\% and 15.2\% WER reduction, respectively. In detail, the proposed neural WPD++ method (ID-14) obtains a higher PESQ score on a small angle and more competing speakers cases by comparing it with the conventional WPD method (ID-7). For example, the PESQ for the angle smaller than 15-degree case could be improved from 2.34 to 2.47. Another example is that ID-14 could increase the PESQ from 2.42 to 2.55 for the three competing speakers' case. These observations illustrate that the proposed neural beamformer could have more capability to reduce interfering speech by using highly correlated neighbouring frames. The purely NN based system (ID-4) does not work well, especially for the WER of ASR performance. This is because the purely NN-based method inevitably introduces some non-linear distortion \cite{du2014robust,xu2020neural} which is harmful to the ASR.

ID-9 estimates the masks for $\bf{\Phi}_{\tilde{\bf{S}}\tilde{\bf{S}}}$ and $\tilde{\bf{R}}$ separately and achieves 2.4\% absolute improvement than ID-8 that uses a single shared mask. It indicates that two different cIRMs for $\bf{\Phi}_{\tilde{\bf{S}}\tilde{\bf{S}}}$ and $\tilde{\bf{R}}$ are essential. ID-11 adds the Mag-MSE loss (defined in Eq. (\ref{eq:mag-mse})) to estimate $\sigma$ and improves the performance by 4.9\% comparing with ID-10. By comparing ID-11 and ID-14, the proposed Combo-loss (defined in Eq. (\ref{eq:combo-loss})) reduces the WER by an absolute 1.82\%. This emphasises the importance of the proper $\bf{\sigma}$ estimation to the proposed neural WPD++. By comparing ID-14 and ID-12, the results show the C-Si-SNR loss function on ``est'' achieves better performance than the Si-SNR loss function. For example, ID-14 could reduce the WER from 13.51\% to 12.04\% and increase the PESQ from 2.67 to 2.76 by comapring to ID-12.
In ID-13, we also extract the $\textbf{cIRM}_{\sigma}$ from ID-14 and multiply it with the first channel speech mixture to get the estimated speech. We observe that $\bf{\sigma}$ could also generate enhanced speech after jointly trained with WPD++. Although ID-13 is worse than the final output of WPD++ (ID-14), ID-13 is better than the purely NN system (ID-4) with higher PESQ (2.50) and lower WER (25.50\%). This is because ID-13 is also a purely NN system but jointly trained with WPD++. Almost all of the purely NN systems could inevitably introduce non-linear distortion which is harmful to the ASR system \cite{du2014robust,xu2020neural}.

Fig.~\ref{fig:spectral-demo} visualizes the spectrograms of the speech mixture, $\bf{\sigma}$, reverberant clean speech, dry clean speech, and the outputs of different systems, respectively\footnote{Listening demos (including real-world recording results) are available at \url{https://nateanl.github.io/2020/11/05/wpd++/}}. All methods have some dereverberation capabilities since the ``reverberation tail'' effect is reduced in the spectrograms. Some distortions could be observed in the purely NN-based method (shown in the white dashed rectangle). More residual noise could be seen in the multi-tap MVDR and WPD spectrograms. 

\section{Conclusions and Future work}
\label{sec:conclusion}
For the simultaneous speech separation and dereverberation, we propose a neural ``WPD++'' beamformer that enhances the beamforming module of the conventional WPD beamformer by adding spatio-temporal correlation. The proposed multi-objective loss function achieves better performance than the Si-SNR loss function in terms of PESQ and WER metrics, which indicates an accurate estimation of $\bf{\sigma}$ is the key to success of WPD or WPD++. The final jointly trained complex-valued mask based WPD++ beamformer achieves relative 16.4\% and 15.2\% WER reductions by comparing with the multi-tap MVDR and the conventional WPD. Compared to the purely NN system, the neural WPD++ beamformer reduces most of the non-linear distortion which is harmful to the ASR. In our future work, we will further improve the neural beamformer and design a better loss function that fits for the dereverberation task.

\bibliographystyle{IEEEbib}
\bibliography{refs}

\end{document}